\documentclass[pre,aps,twocolumn,showpacs]{revtex4}
\usepackage{epsfig}
\usepackage{epsf}

\begin{document}
\newcommand{\D}{\displaystyle}
\def\RE{\mathop{\Re e}}
\def\IM{\mathop{\Im m}}
\def\be{\begin{equation}}
\def\ee{\end{equation}}
\def\ba{\begin{array}}
\def\ea{\end{array}}
%

\pre{{\bf 55}, 6171-6185 (1997)}
\title{Conformal Mapping on Rough Boundaries \\I: 
Applications to harmonic problems}
\author{Damien Vandembroucq and St\'ephane Roux }
\altaffiliation[present address ]{Unit\'e Mixte CNRS/Saint-Gobain, 39 Quai Lucien Lefranc 93303 Aubervilliers FRANCE}
\email{damien.vdb@saint-gobain.com, stephane.roux@saint-gobain.com}

\affiliation
{Laboratoire de Physique et M\'ecanique des 
Milieux H\'et\'erog\`enes,\\
Ecole Sup\'erieure de Physique et de 
Chimie Industrielles,\\
10 rue Vauquelin, 75231 Paris Cedex 05, France}

\begin{abstract}
The aim of this study is to analyze the properties of harmonic fields
in the vicinity of rough boundaries where either a constant potential
or a zero flux is imposed, while a constant field is prescribed at an
infinite distance from this boundary. We introduce a conformal mapping
technique that is tailored to this problem in two dimensions. An
efficient algorithm is introduced to compute the conformal map for
arbitrarily chosen boundaries. Harmonic fields can then simply be read
from the conformal map. We discuss applications to 'equivalent' smooth
interfaces. We study the correlations between the topography and the
field at the surface. Finally we apply the conformal map to the
computation of inhomogeneous harmonic fields such as the derivation of
Green function for localized flux on the surface of a rough boundary.
\end{abstract}
\maketitle



Defining and computing effective properties of heterogeneous media 
is a subject which has been studied for a long time, and for which a 
number of powerful techniques have been developped.  In most cases
however, the heterogeneities are considered to lie in the bulk of the
material.  Another type of inhomogeneity is due to the random geometry of
the surface on which boundary conditions are applied.  This study focusses
on this second type.  We will thus consider {\it homogeneous media} which are
limited by a rough surface or interface.  Our purpose here is to introduce
a very efficient way of solving harmonic problems in two-dimensional systems 
for any geometry of the boundary.

The occurence of rough interfaces in nature is more the general rule
than the exception.  Apart from very specific cases such as mica where
a careful cleavage can produce planar surfaces at the atomic scale,
surfaces are rough.  Even glass with a very homogeneous composition,
where the surface is obtained by a slow cooling of the material, so
that surface tension can act effectively to smoothen all
irregularities, does display roughness in the range 5 to 50 nanometers
over a window of a few micrometers width\cite{Glass}.  Similarly, the
so-called ``mirror'' fracture surface which is optically smooth
exhibits specific topographic patterns when examined with an atomic
force microscope\cite{Guilloteau}.  The key question is thus to
identify the relevant range of scales at which roughness appears.
From common observations, this question may not admit a simple clear
cut answer.  Indeed in a variety of cases, the amplitude of the
roughness appears to be strongly dependent on the size of the examined
surface.  A particular class of such scale dependent roughness, namely
{\it self-affine roughness}\cite{Mandelbrot,Feder}, has recently motivated a
lot of activity (see References
\cite{HalpinPR95,Meakin93PR,BarabasiCUP} for recent reviews) both
because of its relevance in many different instances, and of their
theoretical justification which has been obtained in statistical
physics for a wide class of models ranging from growth
models\cite{Vicsek}, molecular beam epitaxy\cite{Wolf}, fracture
surfaces\cite{fracts}, to immiscible fluid
interfaces\cite{fluidinter}.  Although the present study is not
specific to self-affine surfaces, we shall consider this particular
class in order to apply our method.  The interest of this choice being
that {\it i)} the description of the roughness is realistic for a
number of applications, {\it ii)} consequences can be expressed in
quite general terms as a function of few parameters directly
accessible experimentally, and finally {\it iii)} the most commonly
studied roughness models are ``monochromatic'' surfaces with a single
asperity pattern repeated periodically, and hence the transposition to
more complex geometries may reveal wrong (examples of such cases will
be explicited in the main body of this article).

As previously mentioned, if most surfaces are rough, this roughness may
be of small amplitude macroscopically, and thus one may feel that its role  
can be neglected in most cases.  Fortunately, this is generally true.
Taking into account precisely the surface roughness may be required 
in two distinct classes of problems: 

The first class (I) covers applications where 
the roughness cannot be neglected at the scale at which the bulk 
field varies.  For obvious reasons, there is no way to avoid the 
accurate description of the boundary.  We may mention the following 
potential applications:
\hfill\break$\bullet$  In confined geometries, such as naturally 
encountered in surface force study, the roughness of the surface 
may affect the interpretation and thus the precision of the measurements
since the distance between two facing surfaces is generally estimated from indirect measurements of transport in the gap between the surfaces.\cite{Loubet}
\hfill\break$\bullet$  Field which are rapidly varying in space will be 
sensitive to fine details of the boundary geometry.  The most obvious 
example in this field is the reflection and scattering of a wave by a 
rough boundary.\cite{wave} Of particular importance is the case of 
surface waves, evanescent waves, Rayleigh waves in elasticity,  ...
\hfill\break$\bullet$
In a similar spirit, diffusion processes may display 
anomalous behaviors at short times where the diffusion length is smaller 
or comparable to the roughness.\cite{VBR-EPL95}

The second class of problems (II) where roughness cannot be neglected 
is when one has to focus on the boundary, either because only this
part matters for extraneous reasons or because the system is 
sensitive to high fields which can be induced by the roughness itself.
Some examples of these two cases are listed below:
\hfill\break$\bullet$ Surface phenomenon such as electrofiltration 
requires a proper solution of say a Stokes flow field, in the immediate 
vicinity (typically Debye length scale) of a rough boundary where an 
electric boundary layer is present
and can be entrained by the fluid to give rise to an electric current 
in response to a fluid flow in a porous medium.\cite{electrofiltration}
\hfill\break$\bullet$ The brittle fracture of glass is generally due to 
surface defects which induce locally high stresses, which reduce 
significantly the breaking limit of this material.  In the absence of 
specific surface degradation the most important source of surface defect 
is the topography itself\cite{brittleglass}.  
\hfill\break$\bullet$ Some growth models have a local growth rate 
which depends on a harmonic field locally.   The development of 
unstable modes which will finally induce a macroscopic roughening 
does require the proper analysis of the field at the 
surface.\cite{Feder,Vicsek}

The relative independence of the bulk field on the small scale roughness
of the boundary for a slowly varying field (class II problems) can be used to explore the 
local field close to the boundary using an asymptotic analysis with a 
double scale technique.  The large scale problem consists in solving the 
problem at hand replacing the rough boundary by a smooth equivalent one.
The small scale problem deals with the details of the rough boundary and   
matches at ``infinity''with a homogeneous field.  This local problem will be
considered in full details in the following.

These examples are obviously not exhaustive.   Inhomogeneous boundary 
conditions may arise for instance in contact problems where the roughness
cannot be neglected.\cite{Borodich,RSVH} One may also consider
application outside the realm of physical applications, such as 
the use of harmonic problems and particularly conformal map for
providing a simple means of meshing a domain limited by a rough boundary. 

In the present article we will essentially focus on harmonic problems.
The latter arise in a variety of different domains in physics, such as
electrostatics, thermal or concentration diffusion, flow in porous
media, anti-plane elasticity to mention a few.  Another use of
conformal mappings is the resolution of bi-harmonic problems near a
rough interface; both stress field in elasticity and velocity field in
low Reynolds number fluid mechanics\cite{Hasi,Hig85,Poz87,Pozbook,Poz95}
can be derived from potentials that obey such bi-laplacian
equations. We refer the reader to the companion paper which is
completely devoted to this specific problem.

This paper is devoted to the study of harmonic problems in 2D
semi-infinite media limited by a rough boundary. To extend the
definition of the profile of the boundary to infinity, we use periodic
boundary conditions along the boundary. Although very specific, this
type of geometry will be very convenient as soon as no other boundary
lies close to the first one. The distance threshold to consider in
such a case is typically of the order of magnitude of the larger
spatial wavelength of the profile {\it i.e.} the spatial period in the
geometry we have described.  We use a conformal mapping technique. It
consists in contructing a map from the domain of interest (in the
complex plane) onto a regular semi-plane.  The conformity of the map
allows to preserve harmonicity through the map transform.

In a first part of this paper, we define the form of the conformal mapping 
suited to our geometry.  Then, we adress the question of constructing the 
mapping associated with any prescribed interface. We show that this 
problem can be solved with an iterative algorithm using Fast Fourier 
Transforms (FFT).  This algorithm allows to get the conformal map
in a few iterations of FFT, whose computation time scales as $N\log(N)$ 
where $N$ is the number of Fourier modes used to describe the interface. 
Note the remarquable efficiency of such a technique, considering that the 
map gives the solution of a Laplacian field in the entire two-dimensional 
problem. This problem is very close to the so-called ``Theodorsen problem'' 
\cite{Henrici3} in a circular geometry. We also show that one can generate 
maps which naturally give rise to self-affine boundaries, a powerful 
technique to explore generic properties of such problems.  Specific 
applications of this technique to self-affine profile are studied, 
which includes {\it i)} the question of defining an equivalent smooth 
(planar) interface, and finding its height compared to the geometrical 
average height of the interface, {\it ii)} the correlation between the 
height and the field which is computed exactly in the limit of a small 
roughness amplitude.  These two examples demonstrate the unexpected 
difference in behaviors for persistent and anti-persistent profiles.
Finally, we give the expression of the Green function for localized
flux on a rough interface.

  \section{Transformation conforme}

\subsection{Notations}

\begin{figure}
\epsfxsize 0.8\hsize>
\hfill\epsfbox{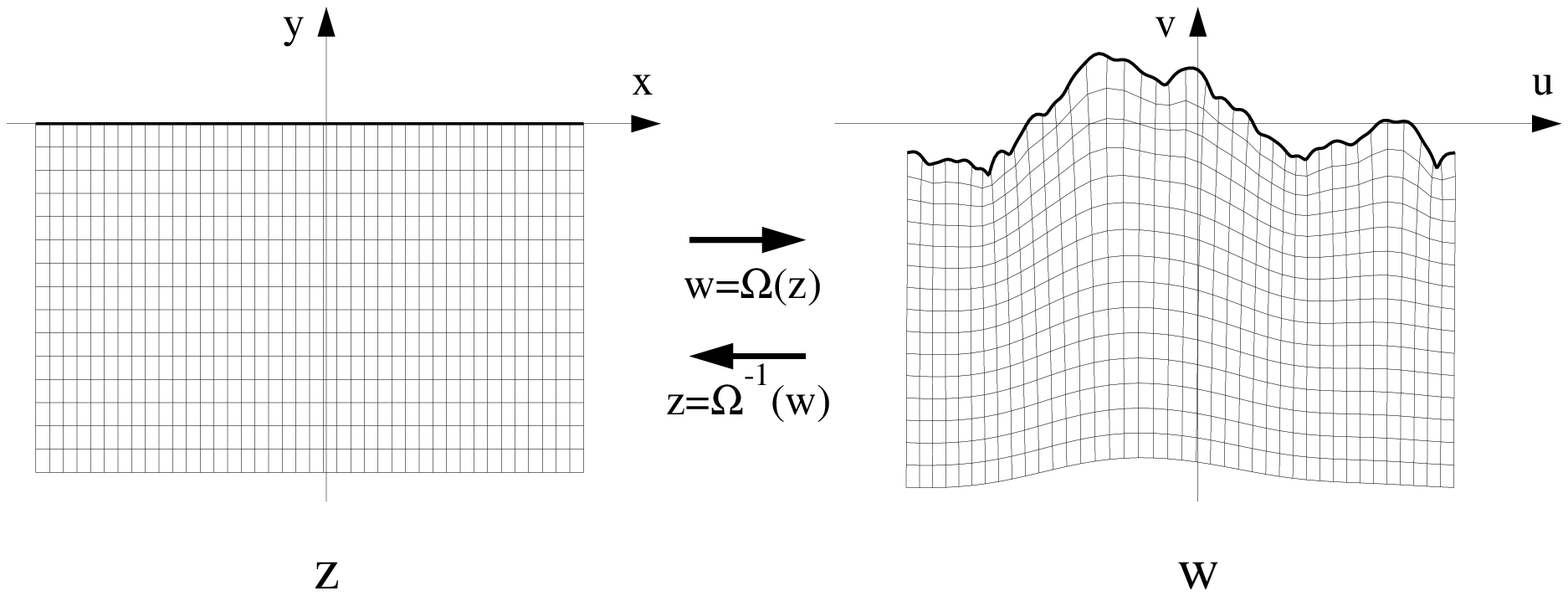}\hfill~
\caption {Illustration de la transformation conforme $\Omega$  introduite dans cette
section.  
A schematic illustration of the mapping $\Omega$ which maps the
semi-infinite plane ${\cal D}$ onto the domain limited by a rough interface 
${\cal E}$. 
\label{fig:method}} 
\end{figure}

In order to study harmonic fields in two dimensions, very powerful
techniques have been developed based on complex analysis\cite{Henrici3}. Among
these, we will use in the following conformal maps, which allow to relate
the geometry we wish to study (i.e. a semi-infinite domain limited by a
rough interface), to a regular one as schematically illustrated in Figure 
(\ref{fig:method}).

As usual, we will identify a point in the plane $(x,y)$, with the complex
number $z=x+iy$. We note $\overline z = x-iy$ the complex conjugate of $z$.
The two variables $z$ and $\overline z$ can be treated as independent
variables instead of $x$ and $y$. A mapping from the complex plane onto
itself is simply defined as a complex function $\Omega$ of $z$ and $%
\overline z$ which transforms one point of the complex plane $z$ into
another point $\Omega(z,\overline z)$. For the mapping to be of physical
interest it has to be bijective in a domain of interest, and thus
inversible. The mapping is {\it conformal} if the function is {\it %
holomorphic}, i.e. it only depends on $z$ and not on $\overline z$. It can
be shown that in this case, local angles are preserved in the transformation
--- apart from singular points --- and hence the term ``conformal''.
Moreover, the real part and the imaginary part of any such holomorphic
function are both harmonic, i.e. $\nabla^2 \mathop{\Re e} \Omega = \nabla^2 %
\mathop{\Im m} \Omega =0$. The latter property results from the expression
of the Laplacian operator in terms of the variables $z$ and $\overline z$: 
\begin{equation}
\nabla^2 \equiv\partial_{xx}^2+\partial_{yy}^2=4\partial_{z\overline z}^2\;.
\end{equation}

Two obvious choices are well suited for our geometry: i) a semi-infinite
plane ${\cal D}$, $\mathop{\Im m}(z)\le 0$ ii) the unit circle ${\cal C}$, $%
\vert z\vert \le 1$. These two domains can be related by the transformation $%
z\to \log(z)$, and thus they are basically equivalent. Since the boundary we
consider is periodic in the $x$-direction, the mapping to the unit circle is
well suited. However, in the following, we will rather use the mapping to
the half-plane, since it corresponds directly to the ``reference'' problem
where the roughness vanishes.

\subsection{Specific transformations}

The domain of interest, denoted by ${\cal E}$ is limited by a rough
interface $\partial {\cal E}$ which is a periodic function of $\mathop{\Re
e}(z)$ of period $X$. The conformal map, $\Omega$, is a function of $z$
which associate one point of the reference domain, ${\cal D}$ or ${\cal C}$,
to another point in ${\cal E}$. From now on in order to distinguish the
initial and the image domain, we will note $w=u+iv$ a point in the image
plane, and keep the notation $z=x+iy$ for the initial plane unless otherwise
mentioned. Before specifying the particular form of the boundary, it is
possible to guess an adequate form for these transformations.

Let us first consider the mapping from the half-plane ${\cal D}$ to ${\cal E}
$. As $\mathop{\Im m}(z)$ tends to $-\infty$, the mapping should approach
the identity, $\Omega(z)\to z$, since the roughness of the boundary is not
expected to play any significant role at a large distance (compared to the
period $X$) from the boundary. We introduce the function $\omega(z)$ such
that 
\begin{equation}
\Omega(z)=z+\omega(z)\;.
\end{equation}
Functions of the form $\exp(-ikz)$ with $k$ real, thus appears to be natural
candidates for $\omega(z)$. They are indeed periodic functions of $%
\mathop{\Re e}(z)$, and vanish exponentially as $\mathop{\Im m}(z)$ goes to $%
-\infty$ when $k>0$. Moreover, in order to satisfy the same periodicity as $%
\partial {\cal E}$ we require that $kX=2n\pi$ where $n$ is an integer. Thus
we propose as a writing of the transformation $\Omega$ the following
decomposition 
\begin{equation}  \label{map}
\Omega(z)=z+ \sum_{n=0}^\infty \omega_k e^{-2i\pi kz/X}\;.
\end{equation}
Without loss of generality we will set $X=2\pi$ for all the rest of this
study.

The rough boundary is to be identified with the image of the $x$ axis, so
that $\partial {\cal E}$ obeys the parametric equations 
\begin{equation}  \label{boundary}
\left\{
\begin{array}{ll}
u=x+ & \mathop{\Re e}\left (\sum_k \omega_k e^{-ikx}\right )\;, \\ 
v= & \mathop{\Im m}\left (\sum_k \omega_k e^{-ikx}\right )\;.
\end{array}
\right .
\end{equation}

The corresponding transformation from the unit disk ${\cal C}$ to the domain 
${\cal E}$ can be obtained from the above form (\ref{map}) and the
tranformation from the disk to the semi-infinite plane $\mathop{\Im m} z \le
0$. The resulting transformation reads 
\begin{equation}  \label{mapcircl}
\Omega(z)=i\log(z)+ \sum_k \omega_k z^{k}\;,
\end{equation}
where $X=1$ has been used. Evidently, the image of the unit circle $%
z=e^{i\theta}$ provides the same parametric form as (\ref{boundary}).

The form of the transformation being imposed, one needs to check that the
transformation is bijective: a point should have a single image, and an
image point a unique parent. This condition does impose some restriction on
the transformation $\Omega$. It can be rephrased simply for the
transformation (\ref{map}) as 
\begin{equation}  \label{inversible}
\left\vert {\frac{d \Omega}{d z}}\right \vert_{z=x+iy} >0\;,
\end{equation}
for all $y<0$. In principle, it is sufficient to impose this condition only
in the strict interior of the domain. If $d\Omega/dz=0$ on the boundary, a
kink may appear at this point. In the following, we will assume that the
interface is smooth at a small scale so that poles are forbidden on the
boundary.

\begin{figure}
\epsfxsize 0.8\hsize
\hfill\epsfbox{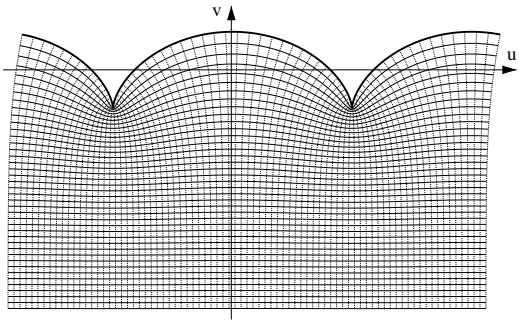}\hfill~
\caption {Image of a regular square grid of the semi-plane ${\cal D}$ using
the transformation (\protect\ref{cycloid}). The parameter $\omega$ has been set to
its maximum value of 1. Cusp appears on the boundary.
\label{fi:cycloid}}
\end{figure}

As an example, if the transformation is simply 
\begin{equation}  \label{cycloid}
\Omega(z)=z+\omega e^{-iz}
\end{equation}
then the condition (\ref{inversible}) reduces to $\vert 1+i\omega e^{ix}
e^{y}\vert >0$ or $\vert\omega\vert<1$. For this maximum value, the image of
the $x$-axis is a cycloid, with cusp points. Figure (\ref{fi:cycloid})
illustrates this limit case.

\section{Computing the mapping for an imposed interface}

The above presented transformation is only useful for a particular
application if the transformation $\Omega$ can be computed, once the
boundary $\partial {\cal E}$ is imposed. This section is devoted to this
problem. The algorithm that we have developed generates the transformation
very efficiently. Different numerical techniques applied to computing the map from arbitrary closed domains to the unit disk can be found in Ref.\cite{Gut83}.
Our algorithm can be shown to be related to the Jacobi method used in these 
studies. 

\subsection{ Description of the algorithm}

We define the rough boundary $\partial {\cal E}$ as a single valued real
function $h$ such that $v=h(u)$ is the equation of the boundary. In other
words, the boundary is given by the parametrised form $w=u+ih(u)$. To comply
with the framework we have chosen here, we use in the following a 2$\pi$
-periodic $h$.

From the particular form of the transformation $\Omega$, we expand the real
and imaginary parts on the boundary $\partial {\cal E}$ 
\begin{equation}
\left\{
\begin{array}{rl}
\label{eqboom} u= & x+\mathop{\Re e}\left[\sum_k\omega_k e^{-ikx}\right] \\ 
h(u)= & \mathop{\Im m}\left[\sum_k \omega_k e^{-ikx}\right]
\end{array}
\right.
\end{equation}
If $u$ were equal to $x$, the second equation would be close to a Fourier
transform expression of the function $h$. More precisely rewriting the last
equation as 
\begin{equation}
h(u)=\mathop{\Re e}\left[\sum_k i\overline{\omega_k} e^{ikx}\right]
\end{equation}
we see that the coefficient $i\overline{\omega_k}$ can be computed from the
Fourier transform of $h(u(x))$. The difficulty is that $u(x)$ is {\it a
priori} unknown. However, we note that if the roughness is small enough, say
of order $\epsilon$, $u$ can be written as $u=x+{\cal O}(\epsilon)$.
Therefore, identifying $x$ with $u$ is a zeroth order approximation. From
the latter, the coefficient $\omega_k$ can be computed by the Fourier
transform of $h(u)$. This provides a first order approximation of $u(x)$,
from which an improved estimate of $\omega_k$ can be obtained, by taking the
Fourier transform of $h(u(x))$, i.e. a non-uniform sampling of the profile $h
$. Iterating this scheme is the basis of our algorithm. We will omit for the
time being the prerequisite on the amplitude of the roughness. We will
return to this point by considering the stability of the algorithm.

The intermediate quantities appearing at the $k$th iteration will be
labelled with a superscript $(k)$. We also formulate the algorithm directly
in discrete terms suited for a numerical implementation. In the remainder of
this article, all functions will be decomposed over a set of $2n$ discrete
values. The number of Fourier modes will thus be limited to $2n$. We first
introduce a series of sampling points $u_j^{(k)}$ with $j=0,...,\ n-1$ which
is initially set to an arithmetic series $u_j^{(0)}=j\pi /n$. The sampling
of $h(u)$ by the $u_j^{(k)}$ gives the array 
\begin{equation}
h_j^{(k)}=h(u_j^{(k)})  \label{eqalgbeg}
\end{equation}
The discrete Fourier transform of this array is the complex valued array 
\begin{equation}
a_j^{(k)}=\sum_{m=-n+1}^nh_m^{(k)}e^{imj}
\end{equation}
for $-n<j\le n$. The latter is shortly written as 
\begin{equation}
a^{(k)}={\cal F}[h^{(k)}]
\end{equation}
where ${\cal F}$ denotes the Fourier transform, which will be chosen as the
Fast Fourier Transform (FFT) algorithm, thus imposing that $n$ is an integer
power of 2. The intermediate mapping $\omega ^{(k)}$ is computed from the $%
a^{(k)}$ as 
\begin{equation}
\left\{ 
\begin{array}{rlcc}
\omega _j^{(k)}= & (i/n)a_j^{(k)} & \qquad {\rm for} & \quad j>0 \\ 
\omega _0^{(k)}= & (i/2n)a_0^{(k)} &  &  \\ 
\omega _j^{(k)}= & 0 & \qquad {\rm for} & \quad j<0 \\ 
&  &  & 
\end{array}
\right. 
\end{equation}
The latter form is obtained from the identification of Eq.(\ref{eqboom}b)
and the definition of $a^{(k)}$, taking care of the fact that one sum is
over positive index, while the other extends over the interval $[1-n,n]$.
Then, one computes the series 
\begin{equation}
\left\{ 
\begin{array}{rlcc}
b_j^{(k)}= & ia_j^{(k)} & \quad {\rm for} &\quad j>0 \\ 
b_0^{(k)}= & 0 &  &  \\ 
b_j^{(k)}= & \overline{b_{-j}^{(k)}}=-i\overline{a_{-j}^{(k)}}=-ia_j^{(k)} & 
\quad {\rm for} &\quad j<0
\end{array}
\right. 
\end{equation}
This linear transformation is shortly noted as 
\begin{equation}
b^{(k)}={\cal G}[a^{(k)}]
\end{equation}
where ${\cal G}$ is the above detailed transformation. The form of ${\cal G}$
is dictated by Eq.(\ref{eqboom}a) for positive index, and from the fact that
the inverse Fourier transform of $b$ (see below) is real. The new sampling
series is finally obtained from 
\begin{equation}
u_j^{(k+1)}={\frac{j\pi }n}+{\cal F}^{-1}[b^{(k)}]  \label{eqalgend}
\end{equation}
The equations (\ref{eqalgbeg}-\ref{eqalgend}) define one step in the
algorithm relating $\omega ^{(k+1)}$ to $\omega ^{(k)}$. Briefly we note
this step $\omega ^{(k+1)}={\cal T}(\omega ^{(k)})$.

The searched function $\Omega$ is clearly a fixed point of the
transformation ${\cal T}$ defined above in a discretized version. The
uniqueness of the transformation $\Omega$ results from that of the harmonic
field in the domain ${\cal E}$ with an equipotential condition on the
boundary and a constant gradient perpendicular to the boundary at infinite
distance from it. Therefore, the only condition to consider is the stability
of the fixed point.

\subsection{Stability}

Let us assume that we have an approximate solution of the
transformation 
$\Omega(z)$, from which we compute the series $u_j$. All intermediate
quantities computed from the exact solution are denoted by a
superscript $*$. Following one complete iteration of the algorithm, we
obtain the following  expressions  
\begin{equation}
\begin{array}{rl}
u_j= & u^*_j+\delta u_j \\ 
h_j= & h^*_j+h^{\prime}(u^*_j)\delta u_j \\ 
a = & a^*+{\cal F}[h-h^*] \\ 
b = & b^*+{\cal G}[a-a^*] \\ 
(\delta u)^{\prime}= & {\cal F}^{-1}[b-b^*]
\end{array}
\end{equation}
where a Taylor expansion of $h$ has been used to estimate the values
of 
$h-h^*$ and where indices are omitted when unnecessary. The resulting
difference $(\delta u)^{\prime}$ after one cycle is thus 
\begin{equation}  \label{eqrecur}
(\delta u)^{\prime}={\cal F}^{-1}\ {\cal G}\ {\cal F}[h^{\prime}(u^*_j)
\delta u_j]
\end{equation}
Let us introduce the norm 
\begin{equation}
\Vert u\Vert^2\equiv\sum_j \vert u_j\vert^2
\end{equation}
Parseval's theorem relates the above norm in real and Fourier spaces
according to 
\begin{equation}
\Vert {\cal F}(h)\Vert=\sqrt{2n} \Vert h\Vert
\end{equation}
In a similar fashion, the transformation ${\cal G}$ does not affect the norm 
\begin{equation}
\Vert {\cal G}(h)\Vert= \Vert h\Vert
\end{equation}
Using the two previous results, we can estimate the norm of $(\delta
u)^{\prime}$ as 
\begin{equation}
\begin{array}{rl}
\Vert (\delta u)^{\prime}\Vert^2 = & \sum_j h^{\prime}(u^*_j)^2(\delta u_j)^2
\\ 
\le & \max(\vert h^{\prime}\vert)^2\Vert (\delta u)\Vert^2
\end{array}
\end{equation}
Therefore, if the absolute value of the slope of the objective profile
satisfies, 
\begin{equation}  \label{eqcondcv}
\vert h^{\prime}(u)\vert < 1
\end{equation}
for all $u$, then the fixed point $\Omega^*$ is attractive for the
transformation ${\cal T}$. It should however be noted that the number of
modes $n$ should be large enough so that the perturbation $\delta u$ should
be small enough to legitimate the Taylor expansion of $h$ used in the
stability analysis.

In practice, the convergence is very fast provided the sufficient condition 
(\ref{eqcondcv}) is fulfilled. Moreover it has to be noted that one step in
the algorithm requires a rather limited amount of computing time of
order 
$n\log(n)$ (i.e. as for a FFT operation). Considering that this computation
gives the solution of an harmonic problem in a semi-infinite domain, this
cost appears to be extremely low.

When our algorithm is applied to a simple monochromatic sine (or cosine)
profile, $h(u)=A\sin(u)$, it turns out that as soon as the condition (\ref
{eqcondcv}) is violated (i.e. $A\ge 1$) the scheme is unstable, and a loop
begins to appear around the origin where the slope exceeds 1. Thus the
sufficient condition is also a necessary condition.

The limit $\vert h^{\prime}(u)\vert < 1$ can simply be broken if one uses 
an under-relaxation scheme.  The optimum determination of the 
under-relaxation parameter, or the use of other algorithms can be found in 
Ref.\cite{Gut83} for mapping arbitrary domains on the unit disk.  The 
transposition of these
algorithms to our problem can be worked out in details.  
Other ways to break this limit is to decompose the transformation $
\Omega$ in two (or more) substeps. Suppose one could map the real axis onto
an intermediate profile using a first transformation $\Omega_1$ and then the
intermediate profile onto the objective one using a second transformation $
\Omega_2$. The combination of the two transformations $\Omega(z)=\Omega_2(
\Omega_1(z))$ is then the searched mapping. By breaking the problem into two
steps, it is possible that each step can be handled by the above presented
algorithm, while the combination of the two giving a profile having a slope
larger than unity. The difficulty here is to device a suited intermediate
step. One could consider for example filtering the initial objective profile
so that the filtered profile may fulfill the slope constraint.  We did not 
investigate this extension any further.

\subsection{Convergence and example application}

We present in the following calculations of conformal transformations
associated with a simple sine interface.

We will use a norm on the error similar to the one introduced in the
previous subsection. The distance $d$ between the objective profile and the
calculated one is defined as follows: 
\begin{equation}
d^2 ={\frac{1}{2\pi}}\int_0^{2\pi} \left[h(u(x))-\Omega(x)\right]^2 dx
\end{equation}
It is convenient to make this distance dimensionless, normalising it by the
amplitude of the profile, $d^{*}=(d /A)$ where the objective profile has the
equation $h(u)=A\sin(u)$.

\begin{figure}
\epsfxsize 0.7\hsize
\hfill\epsfbox{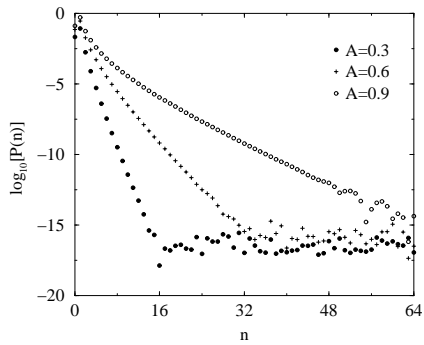}\hfill~
\caption {Power spectrum of the function $\omega$ computed for a sine
profile. The three curves corresponds to three amplitudes, $A=$0.3 
($\bullet$), 0.6 ($+$) and 0.9 ($\circ$).
\label{fi:pssine}}
\end{figure}

It is worth noting that the problem is far from being as simple as it might
appear on the surface. In real space one single Fourier mode is sufficient
to entirely characterise the interface. The transformation $\Omega$ however
requires much more modes. Figure (\ref{fi:pssine}) shows the power spectrum
of the $\omega$ series, for different amplitudes $A$.

\begin{figure}
\epsfxsize 0.7\hsize
\hfill\epsfbox{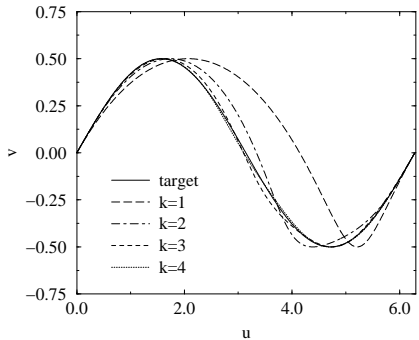}\hfill~
\caption {Images of the real axis obtained after the first $k$ iterates of
the algorithm with the objective profile ($h(u)=A\sin(u)$) shown as a bold
line. In this particular example $A=0.5$ and the number of modes is 32. 
\label{fi:cvsine}}
\end{figure}

The convergence of the algorithm is shown at play on Figure (\ref{fi:cvsine}
), where the obtained profile obtained after the first few iterations are
shown. In this particular example $A=0.5$ and the number of modes is 32.

\begin{figure}
\epsfxsize 0.7\hsize
\hfill\epsfbox{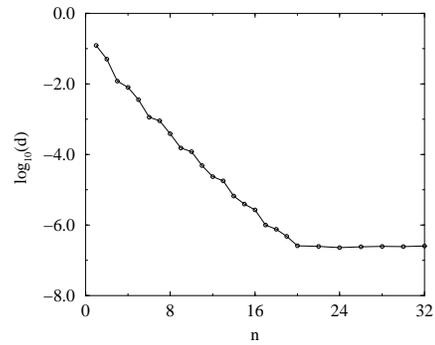}\hfill~
\caption { Minimum error obtained using the algorithm described in the text
for computing the conformal map on a sine profile with amplitude $A=0.5$, as
a function of the number $n$ of modes used in $\omega$.
\label{fi:ersine}}
\end{figure}

The importance of allowing for enough Fourier modes is also illustrated by
considering the minimum error $d$ obtained as a function of $n$ as shown in
Figure (\ref{fi:ersine}). (For this particular study we did not resort to a
FFT algorithm to handle any value of $n$). For $A=0.5$ we observe that about
20 modes are necessary to reach the single precision used in the
computation. As the amplitude increases, the number of modes needed to reach
a small enough error becomes larger and larger.

\section{Self-affine boundaries}

In the description of rough surfaces and interfaces, some recent progress
has been achieved by recognising some scaling invariance properties which
has been observed in a number of real surfaces, and has been shown to result
naturally in a number of growth models. Recent
reviews\cite{HalpinPR95,Meakin93PR,BarabasiCUP} have covered  this field.

Due to the different roles played by the directions normal and parallel to
the surface, the scaling invariance --- when applicable --- involves
different scale factors depending on orientation, a property called
self-affinity. We consider here only two dimensional media so that the
boundary is self-affine if it remains (statistically) invariant under the
transformation 
\begin{equation}
\left\{ 
\begin{array}{rl}
x & \to\lambda x \\ 
y & \to\lambda^\zeta y
\end{array}
\right.
\end{equation}
for all values of $\lambda$. The exponent $\zeta$ is called the ``Hurst'' or
roughness exponent. It is characteristic of the scaling invariance. From
this property, we derive easily that 
\begin{equation}  \label{eqsa}
\langle (y(x)-y(x+\delta))^2\rangle=C^2\delta^{2\zeta}
\end{equation}
where $C$ is a prefactor.

It is noteworthy that the self-affinity property does not involve the
scaling of any measure. However, studying the scaling of the length of the
curve, two regimes are revealed. For large distances, larger than a scale $
\lambda$, the curvilinear length of the profile is simply proportional to
the projected length along the $x$ axis, hence on can identify a trivial
fractal dimension equal to 1. On the other hand, for distances smaller than $
\lambda$, the arc length scales in a non-trivial fashion with the projected
length. This allows to define a fractal dimension equal to $d_f=2-\zeta$.
The cross-over scale $\lambda$ between these two regimes is such that the
typical slope of the profile is 1, i.e. using the notations of 
Eq.(\ref{eqsa}), 
\begin{equation}
\lambda=C^{1/(1-\zeta)}
\end{equation}

Once a roughness profile has been measured, a very convenient way
\cite{Schmittbuhl95PRE} to check
the self-affinity is to compute the Power Spectral Density (PSD) of the
profile. In the case of a self-affine profile of exponent $\zeta$, the PSD
is expected to have the following behaviour: 
\begin{equation}  \label{eqsaps}
P(k) \propto k^{-1-2\zeta}
\end{equation}

It is important to stress that the approach developed in this article is not
specific to self-affine boundaries. However, being given the practical
importance of such boundaries, and the expected generality of scaling
results, we will essentially focus on self-affine boundaries as practical
applications of the concepts developed in the framework of harmonic field in
the vicinity of rough boundaries.

In view of the form of the transformation $\Omega$, and of the previous
scaling, Eq.(\ref{eqsaps}), we introduce a particular set of transformation:
let us choose 
\begin{equation}  \label{eqsyntsa}
\omega_k=A\epsilon_k k^{-1/2 - \zeta}
\end{equation}
where $\epsilon_k$ are random Gaussian variable with 0 mean and unit
variance for the real and imagnary part independently, we can write 
\begin{equation}
\mathop{\Re e} \left[ {\frac{\partial \Omega }{\partial x}} (x+i0) \right] =
1 + A. \mathop{\Im m} \left[ \sum_k \epsilon_k k^{1/2 - \zeta} e^{-ikx}
\right]
\end{equation}
Then for a given set of $\epsilon_k$, we can define a maximum amplitude such
that the mapping is bijective: 
\begin{equation}  \label{eqampmax}
A_{max}={\frac{-1 }{{\mathop{\Im m} \left[ \sum_k \epsilon_k k^{1/2 - \zeta}
e^{-ikx} \right]}}}
\end{equation}
This method gives a short way to generate directly $\Omega$ transforms which
image the real axis to a self-affine interface shape. This approach is
useful to study generic properties of self-affine boundaries.

\begin{figure}
\epsfxsize 0.7\hsize
\hfill\epsfbox{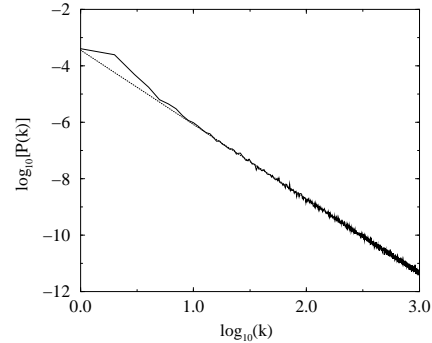}\hfill~
\caption { Power spectrum of the image of the real axis obtained from
synthetic transformation $\Omega $, obtained from
Eq.(\protect\ref{eqsyntsa}). The roughness exponent used is $\zeta
=0.8$. The best power-law regression on this power spectrum has a
slope $s=-1.32$, which corresponds to an estimated roughness
exponent $\zeta'=0.82$, equal within error-bars to $\zeta$. 
\label{fi:pssyntsa}}
\end{figure}

When the amplitude $A$ is small enough, $u\approx x$, and thus the series 
$\omega_k$ is equal to the Fourier transform of the profile. The
transformation $\Omega$ sends the real axis onto a periodic function whose
power-spectrum is of the form Eq.(\ref{eqsaps}). When the amplitude
increases, the first iteration of the algorithm turns out to be rather
approximative. In order to show that the power spectrum of $\omega_k$ is not
significantly altered by further steps, we show in Figure (\ref{fi:pssyntsa}) 
the power spectrum of $\omega_k$ as compared to the initial zeroth order
approximation. The result have been obtained from an average over 100
profiles with 2048 modes each. We see that the synthetic generation of the
transform does not modify the power spectrum of $\omega$ coefficients.

Therefore, we can directly generate mappings $\omega$ which will transform
the real axis into a periodic boundary which is self-affine with any
prescribed roughness exponent for distances smaller than the period. Such a
construction may appear as artificial in the sense that the rough boundary
is not imposed but on the contrary it results from the choice of the
mapping. It is however useful, as will be shown later, because it allows to
study generic properties of harmonic fields close to self-affine boundaries.

The alternative way consists in using the mapping construction algorithm. We
have in a previous section analysed the convergence of the algorithm applied
to the special case of a sine profile. We now consider the case of a
self-affine boundary in a similar fashion. This interface has been
calculated in the real space with 64 modes, and we have used 256 modes in
the conformal transformation. The standard deviation of the height
distribution is called $\sigma$. The chosen $\zeta$ exponent chosen for this
example is $\zeta=0.8$. From Eq.(\ref{eqampmax}), we note that the maximum
amplitude decreases as the number of modes $n$ increases. This is natural
since as the lower cut-off in the scaling regime decreases, the self-affine
function will tend toward a continuous but non-differentiable curve when $%
0<\zeta<1$. The distribution of local slopes is indeed expected to gets
wider and wider as the number of modes increases. Quantitatively, $%
A_{max}\propto n^{\zeta-1}$. It is to be noted that as $n$ increases, the
standard deviation of the height $\sigma$ does not increase. It is to be
 noted that these conclusions are drawn under the hypothesis that the longest
wavelength remains fixed, here set to $2\pi$. Alternatively, if the smaller
cut-off and the amplitude of the corresponding mode were kept constant while
increasing the number of modes, then the maximum amplitude would remain
constant.

\begin{figure}
\epsfxsize 0.7\hsize
\hfill\epsfbox{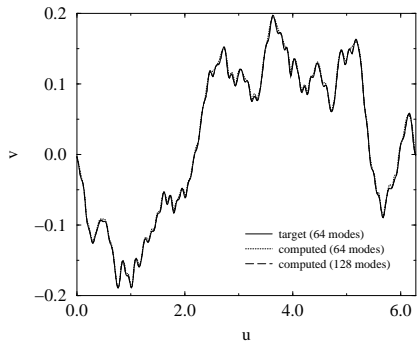}\hfill~
\caption{An example of the obtained profile $\partial{\cal E}$ from the
conformal map, compared to the objective one, chosen to be a self-affine
function with a roughness exponent $\zeta=0.8$. The amplitude of the profile
is 95\% of the maximum amplitude which preserves the convergence of the
algorithm.
\label{fi:exsyntsa}}
\end{figure}

As in the previous example (sine profile), we can observe on Figure (\ref
{fi:exsyntsa}) an example of the conformal map obtained for $%
\sigma/\sigma_{max}=0.95$, where the maximum standard deviation that could
be handled by the algorithm without diverging is $\sigma_{max}\approx 0.1$.
We can see on this figure that the major differences between the objective
and calculated profiles occur in areas where the local slope is maximum. For
roughness amplitude greater than the convergence threshold, one can see
loops appearing in these areas. The convergence speed, the sensitivity to
the number of modes allowed in the determination of $\Omega$, the evolution
of the minimum error, ... behaved for these self-affine profiles in a
similar qualitative way as for the simple sine profile.

\section{Generic properties of harmonic potentials}

In the following, we show that the knowledge of such a conformal
transformation allows to solve immediatly harmonic problems. We essentially
focus here on the case where the field is assumed to be uniform far from the
boundary. This is a typical case as soon as the roughness is of small
amplitude. This can be seen as an asymptotic expansion focusing here on the
small scale details of the interface, whereas the matching with the far
field can be done using a field whose variation is small on the scale of the
roughness amplitude. We will now focus on two problems: perfectly conducting
boundary so that the potential gradient is normal to the boundary, and
perfectly insulating boundary where the potential gradient is parallel to
the surface. Since we know how to taylor mappings which image the real axis
on a generic self-affine boundary, this gives us a key to consider the
scaling features of harmonic fields in the vicinity of self-affine
boundaries.

Harmonic fields are encountered very frequently in nature. Linear transport
involving scalar fields $\Phi$, where the flux $J$ is proportional to the
field gradient plus a conservation law in the absence of sources, and in
steady conditions, $div(J)=0$, imply the harmonic nature of the field $\Phi$%
, $\nabla^2\Phi=0$. Heat diffusion obeying Fourier's law, gives an harmonic
temperature field in steady condition. Mass diffusion with Fick's law is a
similar example with the concentration field. Electric conduction with Ohm's
law, viscous flow in confined two dimensional Hele-Shaw cells, vorticity in
Stokes flow, ... constitutes a partial list of possible applications.

\subsection{Homegeneous far field}

In this part, for the sake of concreteness, we use the case of thermal
conduction. We are interested in the temperature field $T$ in the region $%
{\cal E}$ limited by the rough interface $\partial {\cal E}$. Let us first
consider the case of a perfectly conducting interface, so that $T=T_0$ for
each point of the boundary. We impose in the far field a homogeneous unit
flux of heat. The problem to solve is 
\begin{equation}  \label{eqpblap}
\left\{ 
\begin{array}{rlcc}
\Delta_w T & = 0\quad & {\rm in}\quad & {\cal E} \\ 
T & =T_0 \quad & {\rm on}\quad & \partial{\cal E} \\ 
\vec\nabla T(w) & \to \vec e_v\quad & {\rm if}\quad & v \to - \infty
\end{array}
\right.
\end{equation}

The knowledge of $\Omega$ allows then to define $\Theta$, image field of $T$
in the smooth domain ${\cal D}$: $\Theta (z)= T \circ \Omega(z) = T(w)$. As 
$\Delta_z \Theta = \Delta_w T \left\vert\Omega^{\prime}(z) \right\vert^2$ and 
$\Omega^{\prime}(z) \not= 0$ in ${\cal D}$, the resolution of (\ref{eqpblap}) in ${\cal E}$ is thus
equivalent to: 
\begin{equation}
\left\{ 
\begin{array}{rlcc}
\Delta_z \Theta(z) & = 0\quad & {\rm in}\quad & {\cal D} \\ 
\Theta(z) & =T_0\quad & {\rm on}\quad & {\partial {\cal D}} \\ 
\vec\nabla\Theta(z) & \to \vec e_y \quad & {\rm if}\quad & y \to - \infty
\end{array}
\right.
\end{equation}

The frontier $\partial {\cal D}$ being the real axis, we have immediately
the solution in ${\cal D}$: $\Theta(z)=T_0 +y$ and then the solution in 
${\cal E}$ is: 
\begin{equation}
T(w)=T_0 + \mathop{\Im m}\left[ \Omega^{-1}(w) \right]
\end{equation}

\begin{figure}
\epsfxsize 0.7\hsize
\hfill\epsfbox{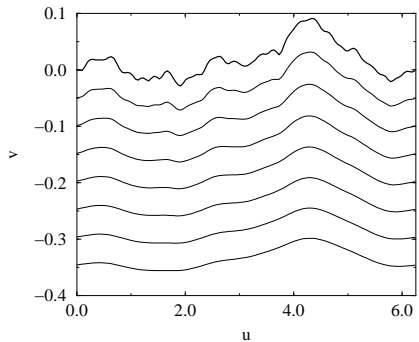}\hfill~
\caption{
Example of isotherm curves close to a rough self-affine boundary on
which the temperature is constant, whereas the temperature gradient is
homogeneous and vertical far from the boundary.
\label{fi:isotherm}}
\end{figure}

Figure (\ref{fi:isotherm}) shows a set of isotherm curves close to a
self-affine isotherm boundary. These lines become smoother and smoother when
the distance to the electrode increases. The morphology of these isotherm
lines have some interesting features. If $\Delta$ denotes the distance to
the boundary, one can observe from the form of the mapping $\Omega$ that
modes with a wavelength smaller than $\Delta$ will be damped whereas longer
wavelength modes will only be slightly decreased. Therefore, the isotherm
curves will be similar to the profile up to a low pass filtering. In the
case of a self-affine boundary, the isotherms will preserve the self-affine
character with the same exponent, but their lower cut-off will increase as
the distance to the actual boundary, up to the distance of order of the
largest wavelength.

Let us now study the temperature gradient. Quite generally, we can write the
gradient in the complex plane as 
\begin{equation}
\nabla_w T(w)\equiv (\partial_u+i\partial_v) T(w)= 2\partial_{\overline w}T
\end{equation}

From the expression of the temperature field, we have 
\begin{equation}  \label{eqgrad}
\left\{
\begin{array}{rl}
\nabla_w T(w) & =i\overline{{\Omega^{-1}}^{\prime}(w)} \\ 
& =i\bigg(1+\overline{\omega^{\prime}(z)}\bigg)^{-1}
\end{array}
\right.
\end{equation}
From the expression of the function $\omega$, we see that at a large
distance from the rough boundary, the term $\omega^{\prime}$ vanishes
exponentially. Therefore, one recovers the imposed condition for the
temperature at infinity i.e. $\nabla T\to i$.

\begin{figure}
\epsfxsize 0.7\hsize
\hfill\epsfbox{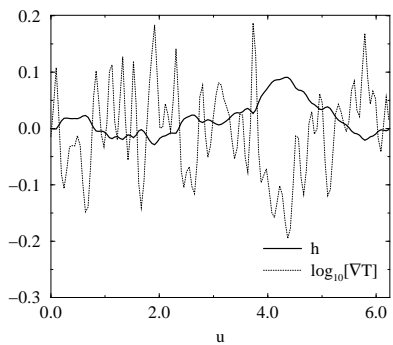}\hfill~
\caption{Profile $h(u)$ of the rough boundary (Symbol $\bullet$) and
temperature gradient $\partial_n T(u)$ on the surface of the same boundary.
We note strong correlations between the two curves.
\label{fi:hgrat}}
\end{figure}

On Figure (\ref{fi:hgrat}) we have presented on the same graph the profile
of a rough electrode and the modulus of the temperature gradient. One may
see quite easily that the field is very large (small) in the deepest
(highest) areas. This field depends naturally both on the local topography
and on its remote environment. The connection between the field and the
local topography can be analysed through cross-correlations as will be done
in a following section.

The perfectly insulating boundary is the other archetypical problem whose
expression is 
\begin{equation}  \label{eqpbdual}
\left\{ 
\begin{array}{rlcc}
\Delta_w T & = 0\quad & {\rm in}\quad & {\cal E} \\ 
\partial_n T & =0 \quad & {\rm on}\quad & \partial{\cal E} \\ 
T(w) & \to u\quad & {\rm if}\quad & v \to - \infty
\end{array}
\right.
\end{equation}
The solution to this problem can simply be obtained from the previous using
duality properties of the harmonic field. The real part of the previous
solution gives the answer to the problem. 
\begin{equation}
T(w)=\mathop{\Re e}\left[ \Omega^{-1}(w) \right]
\end{equation}
The temperature gradient is then simply 
\begin{equation}
\nabla_w T(w)=\bigg(1+\overline{\omega^{\prime}(z)}\bigg)^{-1}
\end{equation}

\subsection{``Equivalent'' smooth boundary condition}

We have seen previously that once we know the conformal mapping able to
transplant the half complex plane $D$ onto the rough domain $E$, we have
immediatly the solution of the electrical potential near the rough
electrode  $\partial E$. If the roughness amplitude remains below the
convergence threshold, we are now able to solve this problem for any
kind of boundary. In practice, very often, one does not worry about
the details of the rough interface. As we have seen most perturbations
die away from the boundary exponentially fast. Therefore, knowing the
longest wavelength of the boundary gives the scale away from the
boundary where the field becomes homogeneous.

This means practically, that if one is interested only in the far field, one
could replace the rough interface by a straight one so that the far field is
unperturbed. The question we want to address in this section is the
following: where should the ``equivalent'' straight interface be located so
as to match the asymptotic far field? A zeroth order guess is to place it at
the geometrical average of the height distribution. This will be shown not
be an exact answer, in the following, we call $H$ the distance between the
equivalent position and the geometrical average.

\begin{figure}
\epsfxsize 0.7\hsize
\hfill\epsfbox{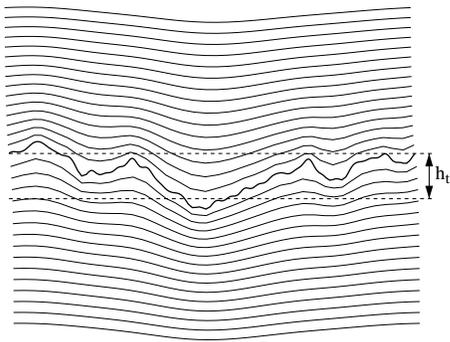}\hfill~
\caption{Schematic illustration of the finite ``electrical'' thickness of a
rough equipotential. The rough equipotential $C$ (shown as a bold curve) is
placed in between two remote planar electrodes. The presence of the rough
equipotential reduces the resistance of the medium in a similar way as an
planar equipotential with a finite thickness. Extrapolating the far
electrical field from the remote electrodes gives ``equivalent'' smooth
boundaries shown as dotted lines. Their relative distance defines the
``electrical'' thickness $h_t$.
\label{fi:elect}}
\end{figure}

In order to illustrate the problem, let us imagine the following experiment.
Let us consider an electrolytic bath, where the electrical field is
homogeneous between two opposite electrodes $A$ and $B$ at a distance $L$
from each other. The electrical resistance $R$ of the set-up is measured.
Then, as illustrated in Figure (\ref{fi:elect}) we place in the middle of
the bath and parallel to the electrodes a rough plane $C$ of negligible
thickness which is a good conductor, so that it can be considered as an
equipotential. We measure again the electrical resistance of the set-up,
which is now reduced to $R-\Delta R$. What is the value of $\Delta R$? So as
to answer this question, we part the system in two, $A-C$ and $B-C$. Each of
these two problems corresponds to the situation described in the
introduction of this section. Extrapolating the field from electrode $A$, we
find an offset $H_1$. Similarly from $B$ we obtain a different offset $H_2$,
so that, ignoring the details of the perturbed field in the vicinity of $C$,
the rough electrode will appear as being equivalent to a plane electrode of
thickness $H_t=H_1+H_2$. This ``electrical thickness'' has nothing to do
with the real thickness of the plane considered here to be zero. If the
rough electrode has the shape of a sine function, of amplitude $A$ and
wavelength $\lambda$ we will argue below that $H_t\propto A^2/\lambda$.
Finally it is a simple matter to relate the resistance drop to this
effective thickness through $\Delta R/R=H_t/L$.

We now revert to the notation of the previous paragraph, and deal with the
temperature instead of the voltage. For distances away from the rough
boundary much greater than $2\pi$, (our longest wavelength), all exponential
terms die out, and hence the far-field can be written 
\begin{equation}
T(w)\approx T_0+\mathop{\Im m}(w-\omega_0)
\end{equation}
where $\omega_0$ is the constant term in the function $\omega$. The off-set
position of the equivalent isotherm is thus 
\begin{equation}
H=\mathop{\Im m}(\omega_0)
\end{equation}

Let us first analyse the problem for a small amplitude sine boundary of
amplitude $A$ and wavelength $\lambda$. The offset $H$ in the location of
the equivalent straight boundary is to be normalised by $A$ to obtain a
dimensionless quantity. The latter should be a function of the dimensionless
ratio $A/\lambda$. Taylor expansion of this function provides the
perturbation expansion 
\begin{equation}
H=a_0 A+a_1 {\frac{A^2}{\lambda}}+a_2 {\frac{A^3}{\lambda^2}}+{\cal O}%
(A^4/\lambda^3)
\end{equation}
A simple argument allows to simplify the latter equation. Suppose one would
analyse the problem for the profile of amplitude $-A$. The latter is
obtained from the former by a translation along the $x$ axis by an amount $%
\lambda/2$. Thus $H$ should be unchanged. This imposes that odd terms in the
expansion should vanish, hence 
\begin{equation}  \label{eqhexpan}
H=a_1 {\frac{A^2}{\lambda}}+{\cal O}(A^4/\lambda^3)
\end{equation}
Thus the dominant correction is of order $A^2/\lambda$. It can be
interpreted as the product of the amplitude $A$ and a typical slope $%
(A/\lambda)$. This result holds in the limit of a small amplitude and long
wavelength. If the wavelength goes to zero, clearly the offset should
converge to the amplitude, but the latter limit cannot be obtained from the
above Taylor expansion in the small parameter $A/\lambda$.

\begin{figure}
\epsfxsize 0.7\hsize
\hfill\epsfbox{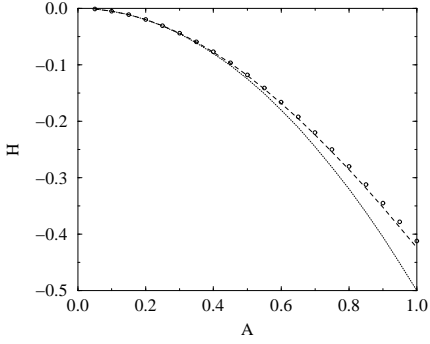}\hfill~
\caption{Offset $H$ of the equivalent boundary from the geometrical average 
of a sine profile of variable amplitude $A$ and fixed wavelength $\lambda 
=2\pi $. The dotted line is the result of the perturbation analysis, 
Eq.(\protect\ref{eqhexpan}) and the symbols are the results obtained from a 
conformal mapping. The dashed curve is the proposed fit Eq.(\protect\ref{eqhconj}). 
\label{fi:eqboh-A}}
\end{figure}

\begin{figure}
\epsfxsize 0.7\hsize
\hfill\epsfbox{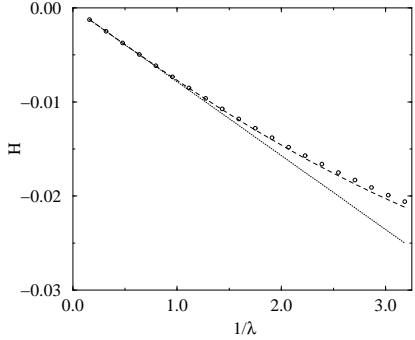}\hfill~
\caption{Offset $H$ of the equivalent boundary from the geometrical 
average of a sine profile of amplitude $A=0.05$ and variable wavelength 
$\lambda $. The dotted line is the result of the perturbation analysis, 
Eq.(\protect\ref{eqhexpan}) and the symbols are the results obtained from 
a conformal mapping. The dashed curve is the proposed fit 
Eq.(\protect\ref{eqhconj}).
\label{fi:eqboh-l}}
\end{figure}

For a sine profile of small amplitude it is possible to carry out the
computation of the coefficient $a_1$. We briefly sketch here the solution.
The potential is to be computed to second order in $A$. We revert as in the
previous sections to a wavelength $\lambda =2\pi $. The solution reads 
\begin{equation}
\begin{array}{llll}
T(w)&=&T_0&\displaystyle +\mathop{\Im m}\left[ w-Ae^{-iw}-i\frac{A^2}{2}
e^{-2iw}\right]
\vspace{0.15cm}
\\
 & & &\displaystyle +\mathop{\Im m}\left[ i\frac{A^2}{2} \right]+ {\cal O}(A^3)
\end{array}
\end{equation}
The offset can be read from this equation as $H=(1/2)A^2$. Reincorporating
the $\lambda $ dependence, we arrive at $a_1=\pi $ or 
\begin{equation}
H=-\pi {\frac{A^2}\lambda }+{\cal O}(A^4/\lambda ^3)
\end{equation}
This last result is of course only valid for small $A$ values, $H$ being
bounded by $A$. On Figures (\ref{fi:eqboh-A}) and (\ref{fi:eqboh-l}), we can
see comparisons between this perturbative calculation and the result
directly obtained by conformal transformation. We observe an excellent
agreement for small amplitudes (resp. large wavelengths) and then the
perturbative calculation overestimates $H$ for larger values of the
amplitude (resp. smaller wavelength). In view of the upper bound on the
offset and the above perturbation expansion, we propose the following form 
\begin{equation}
H\approx -{\frac{2A}\pi }{\rm Arctan}\left( {\frac{\pi ^2A}{2\lambda }}%
\right)   \label{eqhconj}
\end{equation}
which fits the data accurately as can be seen on Figs. (\ref{fi:eqboh-A})
and (\ref{fi:eqboh-l}), and which reproduces both limiting behaviors $%
\lambda \to 0$ and $\lambda \to \infty $.

The question we ask is now how does the result translate to a rough
profile~? In particular for a self-affine profile, there is no
characteristic length 
scales apart from the cut-offs. The product of the amplitude times the slope
is a scale dependent factor. Is it possible to reach quantitative
conclusions for such profiles~?

In order to estimate $H$ for a rough boundary, we use the formalism
developed for introducing the algorithm. We expand the function $\omega $ as
well as all other intermediate quantities in series of the profile
amplitude. Using the linearity of the transformations ${\cal F}$, ${\cal G}$
and ${\cal F}^{(-1)}$, we arrive at 
\begin{equation}
\begin{array}{ll}
H&=\mathop{\Im m}[\omega_0]=\displaystyle{\frac{\mathop{\Re e}[a_0]}{2n}}
={\frac 1{2n}}\sum_jh_j^{*}\\
&\displaystyle ={\frac 1{2n}}\sum_jh^{\prime }(u_j)\ {\cal F}^{(-1)}\circ 
{\cal G}\circ {\cal F}[h_j]
\end{array}
\end{equation}
up to third order terms in the amplitude. We now need an asymmetric version
of Parseval's theorem. Let us compute the integral for two arbitrary arrays
defined in real space for $u$ and Fourier space for $v$ 
\begin{equation}
\begin{array}{rl}
\displaystyle\sum_ju_j{\cal F}^{(-1)}[v]_j= & \displaystyle{\frac 1{2n}}%
\sum_j\sum_ku_jv_ke^{-ikj} \\ 
= & \displaystyle{\frac 1{2n}}\sum_k\overline{{\cal F}[u]}_kv_k
\end{array}
\end{equation}
The offset can now be expressed as 
\begin{equation}
\begin{array}{rl}
\label{eqhexpress}H= & \displaystyle{\frac i{4n^2}}\sum_k\overline{{\cal F}%
[h]}_k\ {\cal G}\circ {\cal F}[h(x)]_k\ k \\ 
= & \displaystyle-{\frac 1{4n^2}}\left( \sum_{k>0}\tilde h_k\overline{\tilde 
h_k}\ k-\sum_{k<0}\overline{\tilde h_k}\tilde h_k\ k\right)  \\ 
= & \displaystyle-{\frac 1{2n^2}}\sum_{k>0}|\tilde h(k)|^2\ k
\end{array}
\end{equation}
where $\tilde h_k$ is the Fourier transform of $h_j$.

\begin{figure}
\epsfxsize 0.7\hsize
\hfill\epsfbox{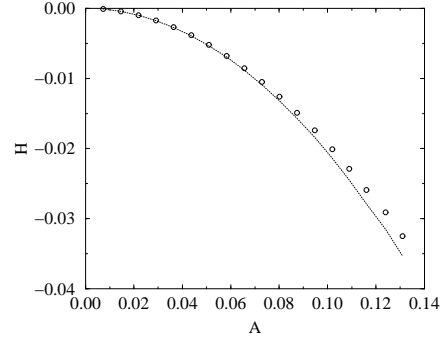}\hfill~
\caption{Offset $H$ of the equivalent boundary from the geometrical
  average of a self-affine profile of variable amplitude $A$, a
  roughness exponent $\zeta =0.8$ and 16 modes. The dotted line is the
  result of the perturbation analysis, Eq.(\protect\ref{eqhexpress}) and the
  symbols are the results obtained from a conformal mapping.
\label{fi:eqbosa}}
\end{figure}

Figure (\ref{fi:eqbosa}) gives the evolution of $H$ with the amplitude of
self-affine profiles of roughness exponent $\zeta =0.8$, and 16 modes.
Again, we observe that the above given expression (\ref{eqhexpress}) is
accurate for small amplitude, but shows deviations for larger amplitudes.

It is interesting to consider the scaling of $H$ observed from the generic
transformations where $\omega _k$ are postulated to be $\omega _k=A\epsilon
_kk^{-1/2-\zeta }$. The expectation value of the offset $\langle H\rangle $
reads to dominant order in the amplitude 
\begin{equation}
\langle H\rangle =A^2\sum_{j=1}^nj^{-2\zeta }  \label{eqhsasca}
\end{equation}
where the extra factor of 2 comes from the expectation value of $\langle
|\epsilon |^2\rangle =2$, since real and imaginary parts of $\epsilon _k$
are independent Gaussian variables of zero mean and unit variance. Depending
on the value of the roughness exponent $\zeta $ two cases are to be
distinguished.

\noindent $\bullet $ For a {\em ``persistent''} profile --- i.e. $\zeta >0.5$
--- the sum in Eq.(\ref{eqhsasca}) is dominated by the smallest $j$, i.e.
the longest wavelength, and thus the scaling of $\langle H\rangle $ can be
expressed as 
\begin{equation}
\langle H\rangle =Z(2\zeta )\frac{A^2}4\left( \frac{2\pi }{{\lambda _{max}}}%
\right) ^{^{-2\zeta }}
\end{equation}
where $Z(s)=\sum_1^\infty k^{-s}$ is the Riemann zeta function. We have
quitted momentarily the convention that the largest wavelength is $2\pi $,
hence $\lambda _{min}$ and $\lambda _{max}$ are respectively the smallest
and largest cut-off lengths in the profile. In this case, $A$, the
amplitude, is such that the largest wavelength mode amounts to $A\epsilon
_1(2\pi /\lambda _{max})^{-1/2-\zeta }$. Let us introduce the standard
deviation of the profile given by 
\begin{equation}
\sigma ^2=(1/\lambda _{max})\int h(x)^2dx=1/(2n)\sum_jh_j^2
\end{equation}
which leads (using Parseval's theorem) to 
\begin{equation}
\langle \sigma ^2\rangle ={\frac 1{4n^2}}\sum_k\left| \tilde h_k\right|
^2=Z(2\zeta +1)\frac{A^2}2\left( \frac{2\pi }{{\lambda _{max}}}\right)
^{-2\zeta -1}
\end{equation}
$\zeta $similar to the scaling Eq.(\ref{eqsa}).

Eq.(\ref{eqhsasca}) can then be expressed as 
\begin{equation}
\langle H\rangle =\pi \frac{Z(2\zeta )}{Z(2\zeta +1)}{\frac{\langle \sigma
^2\rangle }{\lambda _{max}}}  \label{eqpers}
\end{equation}
The latter equation simply means that the rough profile behaves as a the
simple monochromatic profile. This conclusion is however not always valid as
is shown in the following case.

\noindent $\bullet $ For a {\em ``anti-persistent''} profile --- i.e. $\zeta
<0.5$ --- the sum in Eq.(\ref{eqhsasca}) is dominated by the largest $j$,
i.e. the shortest wavelength, in contrast to the previous persistent case. 
\begin{equation}
\langle H\rangle \propto A^2\lambda _{\max }^{2\zeta }\left( {\frac{\lambda
_{min}}{\lambda _{max}}}\right)^{2\zeta -1}
\end{equation}
Therefore, we can express the scaling of $H$ in an intrinsic fashion as 
\begin{equation}
\langle H\rangle \propto {\frac{\langle \sigma ^2\rangle }{\lambda _{max}}}%
\left( {\frac{\lambda _{min}}{\lambda _{max}}}\right) ^{2\zeta -1}
\end{equation}
In contrast to the persistent case, it appears that the offset $H$ is
dependent on the lower cut-off scale of the profile. In fact if $\lambda
_{min}$ is kept fixed, the standard deviation $\sigma $ grows as $\lambda
_{max}^{2\zeta }$. Therefore, one sees that the upper scale cut-off
disappears, so that $H$ only depends on $\lambda _{min}$. In order to see
this more clearly, we introduce another measure of the roughness which is
sensitive to the small scale. Let $\varsigma $ be the norm of the derivative
of $h$: 
\begin{equation}
\varsigma ^2=(1/\lambda _{max})\int h^{\prime }(u)^2du=(1/2n)\sum_jh^{\prime
}(u_j)^2
\end{equation}
which amounts to 
\begin{equation}
\langle \varsigma ^2\rangle \propto A^2\lambda _{\max }^{2\zeta -1}\left( {%
\frac{\lambda _{min}}{\lambda _{max}}}\right) ^{2\zeta -2}
\end{equation}
From the latter norm, the offset can be written as 
\begin{equation}
\langle H\rangle \propto {\langle \varsigma ^2\rangle }\lambda _{min}
\end{equation}
which is the counterpart of Eq.(\ref{eqpers}) for the antipersistent case.

As a conclusion, the scaling of the offset $H$ is controlled by the shortest
(resp. longest) scale cut-off of the self-affine regime for anti-persistent
(resp. persistent) boundaries.

\subsection{Correlation between local field and topography}

In a preceeding section, we have extracted the expression of the temperature
gradient as a function of the transformation $\omega$. We now use it to
investigate the correlations between the topography and the temperature
gradient. We study these correlations in the limit of a small amplitude.

The first order perturbation in the temperature gradient can be extracted
from Eq.(\ref{eqgrad}) as 
\begin{equation}
\vert\nabla T\vert^2=1-\omega^{\prime}(x)-\overline{\omega^{\prime}(x)}+%
{\cal O}(\epsilon^2)
\end{equation}
where the amplitude of the profile is asssumed to be of order $\epsilon$. We
introduce the logarithm of the temperature gradient denoted $\varphi$ which
can be expressed as 
\begin{equation}
\varphi\equiv\log(\vert\nabla T\vert^2) =-2\mathop{\Re e}[\omega'(x)]+{\cal O%
}(\epsilon^2)
\end{equation}
From now on we will omit the ${\cal O}(\epsilon^2)$ term, keeping in mind
that we focus here only on the dominant term.

In order to compute the correlation between the gradient of temperature and
the height, we form the cross-product and average over $u$ (or $x$ for
convenience, since their difference is of order $\epsilon$). The expectation
value of the product is 
\begin{equation}
\langle \varphi(u) h(u)\rangle =-2\langle \mathop{\Re e}[\omega'(x)]%
\mathop{\Im m}[\omega(x)]\rangle =-2 \sum_k k\vert\omega_k\vert^2
\end{equation}
It is amazing that the same expression appeared when computing the offset of
the equivalent straight boundary.

We define now the correlation coefficient $\alpha$ which can be identified
as the slope of a linear regression between $h$ and $\varphi$. Its value is 
\begin{equation}
\alpha\equiv{\frac{\langle \varphi(u) h(u)\rangle}{\langle h^2(u)\rangle}}
\end{equation}
since $\langle h\rangle=\langle \varphi\rangle=0$ to first order in $\epsilon
$. Hence we have 
\begin{equation}
\alpha=-2 {\frac{\sum_k k\vert\omega_k\vert^2 }{\sum_k \vert\omega_k\vert^2}}
\end{equation}
This expression holds for any rough boundary of small amplitude.

In the particular case of a self-affine boundary, we assume as in the
previous section that the transformation $\omega$ can be taken as the one
generated artificially from its Fourier decomposition. The latter expression
can thus be written as 
\begin{equation}  \label{eqalpha}
\alpha=-2 {\frac{\sum_k k^{-2\zeta} }{\sum_k k^{-1-2\zeta}}}
\end{equation}
From the latter expression, we have to distinguish between persistent $%
\zeta>1/2$ and anti-persistent $\zeta<1/2$ profiles depending on whether the
series is convergent or divergent when the number of modes increases.

\noindent $\bullet $ In the case of a {\em persistent} self-affine boundary,
as the number of modes increases to infinity, the value of $\alpha $
converges toward an asymptotic limit $\alpha ^{*}$ given by a ratio of
Riemann Zeta functions 
\begin{equation}
\alpha ^{*}=-2{\frac{Z(2\zeta )}{Z(1+2\zeta )}}  \label{eqriemann}
\end{equation}
The divergence of the Zeta function as its argument approaches 1, leads to a
divergence of $\alpha ^{*}$ as $(\zeta -1/2)^{-1}$. In the more general case
where $\lambda _{max}$ is not set to $2\pi $, the above equation should be
corrected to 
\begin{equation}
\alpha ^{*}=-2{\frac{Z(2\zeta )}{Z(1+2\zeta )}}\left( {\frac{2\pi }{\lambda
_{max}}}\right) 
\end{equation}

\begin{figure}
\epsfxsize 0.7\hsize
\hfill\epsfbox{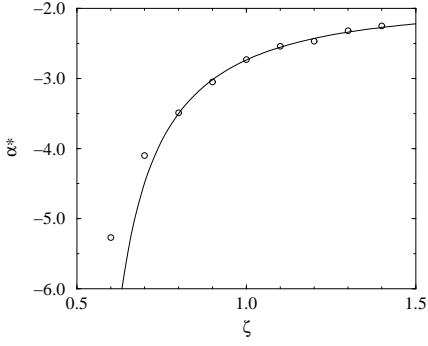}\hfill~
\caption {Average of the logarithm of the temperature gradient $\langle \varphi \rangle =\langle \log (|\nabla T|^2)\rangle $, for fixed height $h$, as a function of $h$. The data points (symbols $\bullet $) are averages over 1000 profiles of small amplitude, with a roughness exponent $\zeta =0.8$, and 64 Fourier modes. The theoretical prediction is shown as a dotted line of slope $\alpha ^{*}$.
\label{fi:corrstat}} 
\end{figure}

\begin{figure}
\epsfxsize 0.7\hsize
\hfill\epsfbox{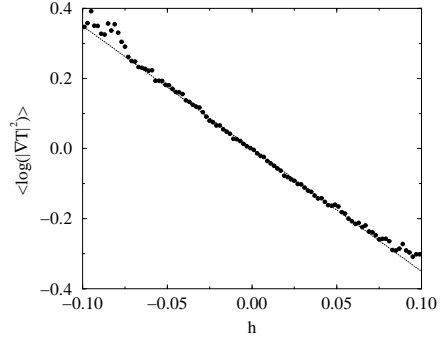}\hfill~
\caption{Evolution of the limit $\alpha^{*}$ as a function of the roughness exponent $\zeta $. As $\zeta $ approaches 1/2, the coefficient diverges as $(\zeta -1/2)^{-1}$. The dotted line shows the predicted behavior and the symbols are the results computed from conformal mappings.
\label{fi:corrslop}}
\end{figure}

As a practical illustration of the latter property we have studied the
correlations between $h$ and $\varphi $ by averaging $\langle \varphi
\rangle $ at fixed $h$ for 1000 profiles having the same characteristics:
amplitude $A=0.25A_{\max }$, roughness exponent $\zeta =0.8$ and 64 Fourier
modes. Figure (\ref{fi:corrstat}) shows the evolution of $\varphi $ versus $h
$. From the Eq.(\ref{eqriemann}) we estimate $\alpha
^{*}=-2Z(1.6)/Z(2.6)\approx -3.50$. As shown on Figure (\ref{fi:corrstat}),
this value of $\alpha ^{*}$ provides an accurate  fit to the data. The
evolution of this coefficient as a function of $\zeta $ is shown in Figure
(\ref{fi:corrslop}).

\noindent$\bullet $ The anti-persistent self-affine profile behaves
differently from the previous case. The correlation between the surface
temperature gradient and the height vanishes. Mathematically, this result
can be traced to the difference of behavior of the two series in Eq.(\ref
{eqalpha}). However, as in the previous section concerning the location of
the equivalent smooth interface, one can extract the asymptotic behavior of 
$\alpha $. 
\begin{equation}
\label{eqalpha_ap}
\begin{array}{lll}
\alpha &= &\displaystyle{ 
\frac{\langle \varphi(u)h(u)\rangle }{\langle h'^2(u)\rangle }
\frac{\langle h^{\prime}(u)^2\rangle }{\langle h(u)^2\rangle } }
\vspace{0.15cm}
\\
&= &\displaystyle{ 
-2\left( \frac{2-2\zeta }{1-2\zeta }\right) 
\left( \frac{\lambda_{min}}{2\pi }\right) 
\left( \frac{\varsigma ^2}{\sigma ^2}\right) }
\end{array}
\end{equation}

This latter result shed some light on the physical meaning of the previously
mentioned divergence. In our presentation, we have chosen to fix the largest
wavelength (set to $\lambda_{max}=2\pi$) and amplitude of this mode.
Increasing the number of modes implies that the shortest wavelength $%
\lambda_{min}$ decreases. For roughness exponents in the range $0<\zeta<1$,
this implies an algebraic increase of $\varsigma^2$ with $n$, while $\sigma^2
$ is bounded. This divergence of $\varsigma^2$ is of no importance for the
correlation $\alpha$ only if the profile is persistent. Otherwise, Eq.(\ref
{eqalpha_ap}) holds. The perturbation method used however assumes that both $%
\sigma^2$ and $\varsigma^2$ should be small. The above analysis simply
identifies which cut-off will dictate its behavior to the correlation. The
antipersistent case is more suited to the case where $\lambda_{min}$ is
fixed together with its amplitude, while $\lambda_{max}$ varies. In this
case, the $\alpha$ coefficient increases as $\lambda_{max}^{1-2\alpha}$ as
can be read from Eq.(\ref{eqalpha_ap}) using the scaling $\sigma_2\propto
\lambda_{max}^{2\zeta}$ ($\varsigma$ being independent of $\lambda_{max}$).

\subsection{Green function for harmonic problems on a rough interface}

Up to now, we have only considered harmonic problems with a uniform field at
infinity. This kind of boundary condition is of particular interest for
problems where the scale of variation of the field in the bulk of the solid
is large compared to the scale of the roughness so that an asymptotic
development can be performed where the matching is to be done on the far
field as one focusses on the rough boundary. However, from the conformal
mapping, one can address more complex types of boundary conditions.

In order to illustrate this, we develop here a particular class of solutions
which can be used to solve any problem. We will consider Green functions
which give the field in the medium for localized flux $f$ injected in the
medium from the surface.

Let us consider the following problem: a localised flux $f=1$ is injected at
point $(1,0)$, on the border of the unit circle ${\cal C}$. The remaining
boundary is perfectly insulating. The same flux is withdrawn at the origin
where $f=-1$. The harmonic field which fulfills such boundary conditions is 
\begin{equation}
\Phi(z)=-\mathop{\Re e}\left[\log\left({\frac{(z-1)^2}{z}}\right)\right]
\end{equation}
This potential $\Phi$ is the Green function for the Domain ${\cal C}$.
Considering the transformation $z\to -i\log(z)$ maps the unit circle to the
semi-plane ${\cal D}$. In the transformation, the potential $\Phi$ becomes 
\begin{equation}
\Phi(z)=-\mathop{\Re e}\left[iz+2\log\left(1-e^{-iz}\right)\right]
\end{equation}
which is the Green function for a unit flux localised at every site $%
(2k\pi,0)$ for all integer $k$. At infinity, the potential approaches $%
\Phi(z)\to y$. From this Green function it is simple to derive the one
obtained for a translated array of sources. For sources at $(x_0+2k\pi,0)$,
we have 
\begin{equation}
\Phi(z,x_0)=-\mathop{\Re e}\left[iz+2\log\left(1-e^{i(x_0-z)}\right)\right]
\end{equation}

From this latter expression, the Green function for a localised and periodic
source on the rough profile is obtained by combining $\Phi$ and $\Omega$.
The Green function thus reads 
\begin{equation}
\Psi(w,w_0)=\Phi\left(\Omega^{-1}(w),\Omega^{-1}(w_0)\right)
\end{equation}
which gives the potential at point $w$ for a series of sources periodically
spaced with the same period as the profile $w_0+2k\pi$.

\section{Conclusion}

We have introduced here conformal mapping technique which allows to address
harmonic problems in semi-infinite domains limited by a rough interface.
This mapping is accompanied by an efficient numerical technique which allows
us to compute the mapping by a few iterations of a one dimensional Fourier
transform. Moreover, this technique provides a natural basis for discussing
analytically some practical applications.

We then defined and studied the notion of an equivalent smooth boundary,
whose position has been obtained exactly in the limit of a small amplitude.
This question underlined the differences between persistent and
anti-persistent boundaries, in terms of sensitivity to the lower or upper
scale cut-off of the self-affine character of the boundary.

We considered the question of correlations between the gradient of the
harmonic field on the boundary and the height of the profile at the same
point. The correlation has been explicitly computed and shown to converge to
a precise limit for persistent boundaries. Anti-persistent profiles lead to
a correlation coefficient which is dependent on the self-affinity range.

Finally we have shown that the same mapping can be used to address different
boundary conditions, including the extreme case of a point like source on
the boundary which is treated exactly.

Extensions of the above technique are numerous. We essentially focused here
on static problems involving harmonic fields. However, the same mapping may
also be used in connection with evolution problems such as diffusion or wave
propagation (localisation). Thermal diffusion in the vicinity of a rough
boundary has recently been shown to display an anomalous scaling behavior at
early stages which could be addressed by such methods. The a.c. impedence of
rough electrodes is another potential field of extension which has been
studied in recent years.

It may also be one constitutive brick of a different mapping dealing with
different geometries. An example of such extensions is the stress intensity
factor (i.e. the singular behavior of the stress field) at a crack tip. In
the framework of antiplane elasticity one can compute the local stress
intensity factor at the crak tip and relate it to the far-field singular
behavior. This problem is currently being investigated.

In a companion paper we extend our mapping to computations of bi-harmonic
fields with applications to Stokes flow close to rough boundaries and
elastic stress fields close to a rough surface.


\end{document}